\documentclass[11pt]{article}
\usepackage{moriond,epsfig}
\usepackage[latin1]{inputenc}
\usepackage[OT1]{fontenc}
\usepackage{subfig}

\def\Journal#1#2#3#4{{#1} {\bf #2}, #3 (#4)}

\def\AA{\em A\&A}
\def\APJ{\em ApJ}

\begin{document}
\vspace*{4cm}
\title{Dark Energy Survey Supernovae: Simulations and Survey Strategy}

\author{J.~P. Bernstein}

\address{Argonne National Laboratory, HEP Division, Argonne, IL 60439}

\author{R. Kessler}

\address{University of Chicago, KICP, Chicago, IL 60637}

\author{S. Kuhlmann and H. Spinka}

\address{Argonne National Laboratory, HEP Division, Argonne, IL 60439}

\author{For the Dark Energy Survey Collaboration}

\address{}

\maketitle\abstracts{
We present simulations for the Dark Energy Survey (DES) using a new code
suite (SNANA) that generates realistic supernova light curves
accounting for atmospheric seeing conditions and
intrinsic supernova luminosity variations using MLCS2k2 or SALT2 models. 
Errors include stat-noise from photo-statistics and sky noise. 
We applied SNANA to simulate DES supernova observations and employed an
MLCS-based fitter to obtain the distance modulus for each simulated
light curve. We harnessed the light curves in order to study selection
biases for high-redshift supernovae and to constrain the optimal DES
observing strategy using the Dark Energy Task Force figure of merit.
}

\section{Introduction}
The Dark Energy Survey (DES) is on track for first light in 2011 and will carry out a deep optical and near-infrared survey of 
5000 square degrees of the South Galactic Cap to $\sim$24th magnitude using a new 3 square-degree CCD camera (called DECam) to be mounted on the Blanco 4-meter telescope at CTIO. DES uses thicker CCDs from Lawrence Berkeley National Laboratory with greater red sensitivity as compared to previous surveys. In exchange for the camera, CTIO will provide DES with 525 nights on the Blanco spread over 5 years. 
The survey data will allow the measurement of the dark energy and dark matter
densities and the dark energy equation of state through four independent methods: galaxy clusters, 
weak gravitational lensing tomography, galaxy angular clustering, and supernova (SN) distances. While the logistics of the SNe survey are still being finalized, time allocation within the larger survey will be $\sim$1000 hrs (yet to be finalized) with maximal use of non-photometric time (up to 500 hrs). Likewise, the spectroscopic follow-up strategy is still being fleshed out. The working estimate is currently 25\% with the remaining redshifts to be obtained via host-galaxy follow-up.

The DES SN working group has undertaken simulations of DES 
observations with the goal of constraining the optimal SN survey strategy. Toward this end, 
we apply the SN simulation package (SNANA) developed by Kessler for the SDSS-II SN 
Survey and later modified for non-SDSS surveys. SNANA generates 
realistic light curves accounting for atmospheric seeing conditions, host-galaxy extinction, cadence, 
and intrinsic SN luminosity variations using MLCS2k2 (Jha {\em et al.} 2007 [1]) or SALT2 (Guy {\em et al.} 2007 [2]) models. The simulation errors include stat-noise from photo-statistics and sky noise. The package includes a light-curve 
fitter that shares many software tools, uses the MLCS2k2 model, with 
improvements, and fits in flux rather than magnitudes. In this paper, we present light curve simulations for DES and describe a high-redshift bias that arises when selection effects are not accounted for in the analysis.

\section{The SNANA Package}\label{subsec:snana}
SNANA uses a mixture of C and FORTRAN routines to simulate and fit SN light curves for a range of redshifts (\textit{z}). SNANA generates
fitted distance moduli, $\mu$, and passes $\mu$--\textit{z} pairs to a cosmology fitter. It is publicly available \footnote{http://www.hep.anl.gov/des/snana\_package} and requires CFITSIO and CERNLIB. The simulation is designed to be fast, generating a few dozen light curves per second, while still providing accurate and realistic SN light curves. Using the package requires the generation of a survey library that includes the survey characteristics (e.g., the observing cadence, seeing conditions, and CCD properties). Generating this library is easy post-survey; predicting it before the survey is crucial to making realistic predictions for the light-curve quality. The light curve fitter takes longer to run, up to many hours depending on the number of SNe and number of fit parameters.

\begin{figure}%
\subfloat[An SNANA light curve for redshift \textit{z}$\sim$0.27.]{\includegraphics[angle=0,scale=0.4]{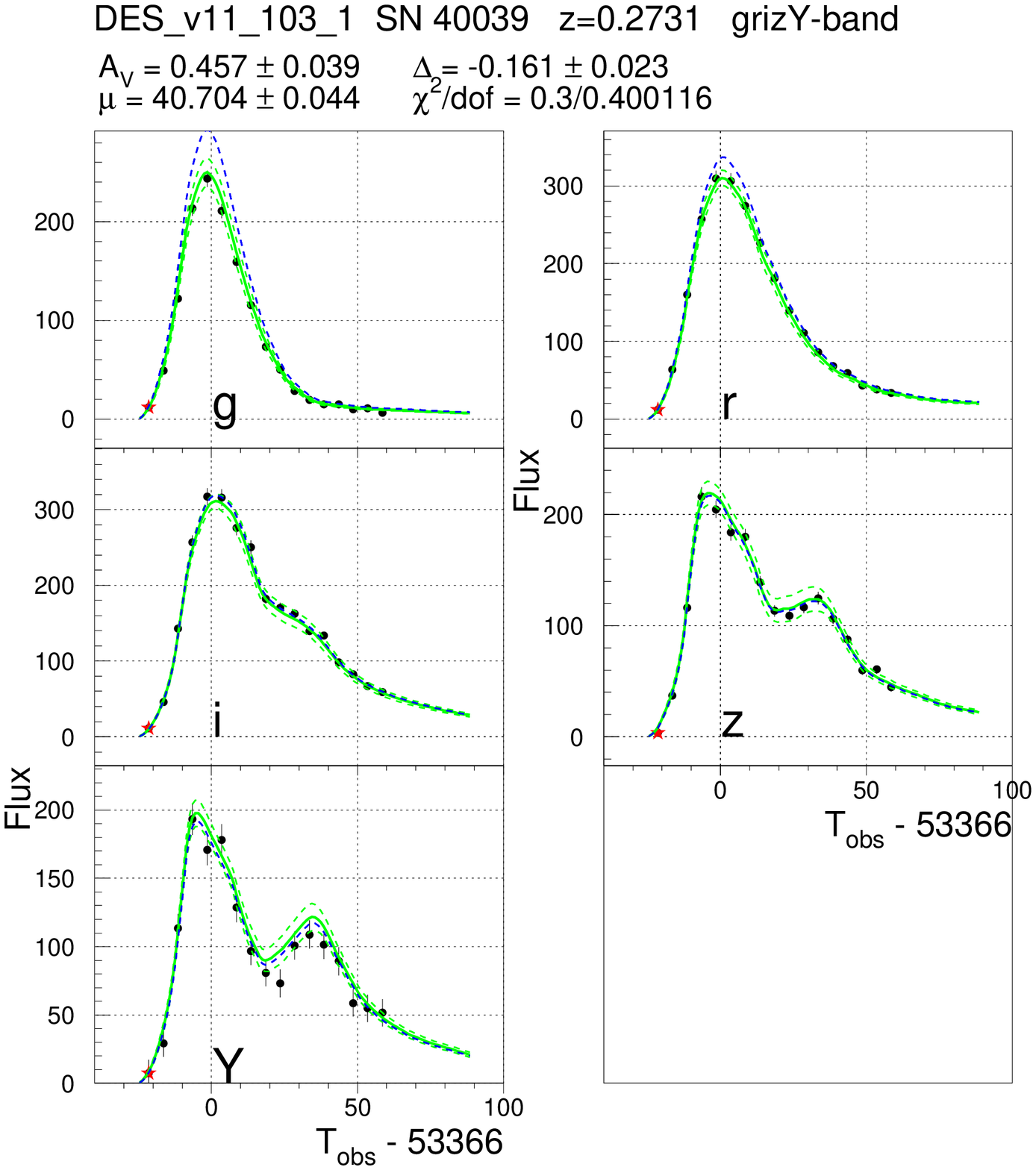}}\hfill
\subfloat[An SNANA light curve for redshift \textit{z}$\sim$0.75.]{\includegraphics[angle=0,scale=0.4]{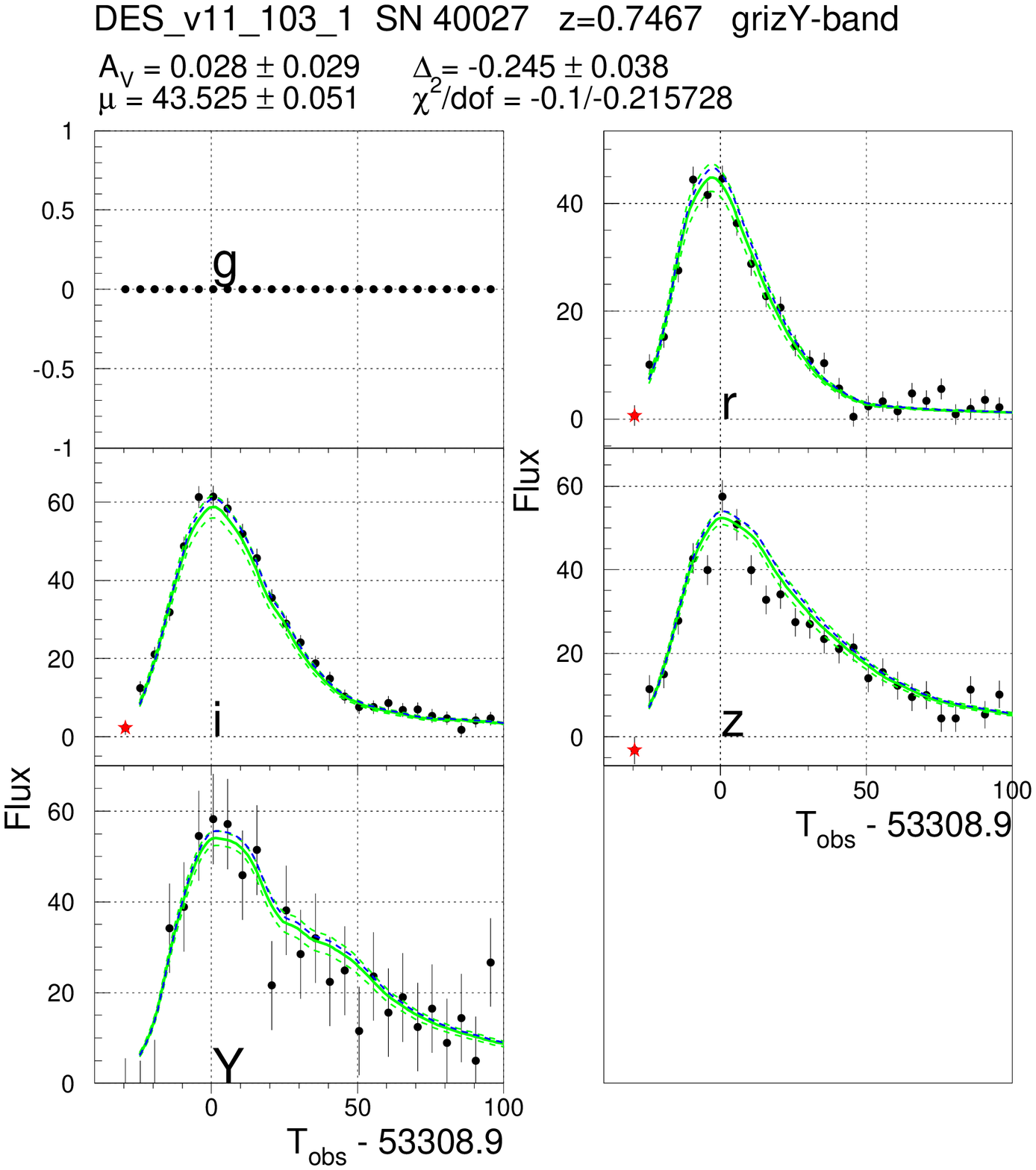}}\\[-10pt]
\captionsetup[figure]{margin=10pt}%
\caption{Plotted is flux vs. time in days. Points are the simulated data, the blue dashed line is with no extinction or fluctuations, and the solid and dashed green lines are the best realistic fit and error bounds. The ``red bump'' at $\sim$40 days is characteristic of SNe Ia, is clearly evident in the Y band for \textit{z}$\sim$0.27, and fades for \textit{z}$\sim$0.75.}
\label{fig:lc}
\end{figure}

\section{Simulations}\label{sec:sims}
For the simulations presented here, we employed the MLCS2k2 model as the basis for generating and fitting SNe light curves. The free parameters are the epoch of maximum light in the B-band ($t_{\rm o}$), the distance modulus ($\mu$), the luminosity/light curve shape parameter ($\Delta$), and the extinction in magnitudes by dust in the host galaxy (parameterized by A$_{\rm V}$ and R$_{\rm V}$ from Cardelli {\em et al.} 1989 [3]). Note that for this work we fixed R$_{\rm V}$ = 3.1 (the average for the Milky Way) but will explore fitting R$_{\rm V}$ in the near future. Fig.~\ref{fig:lc} shows example light curves.

\begin{figure}
\begin{center}
\includegraphics[angle=0,scale=0.715]{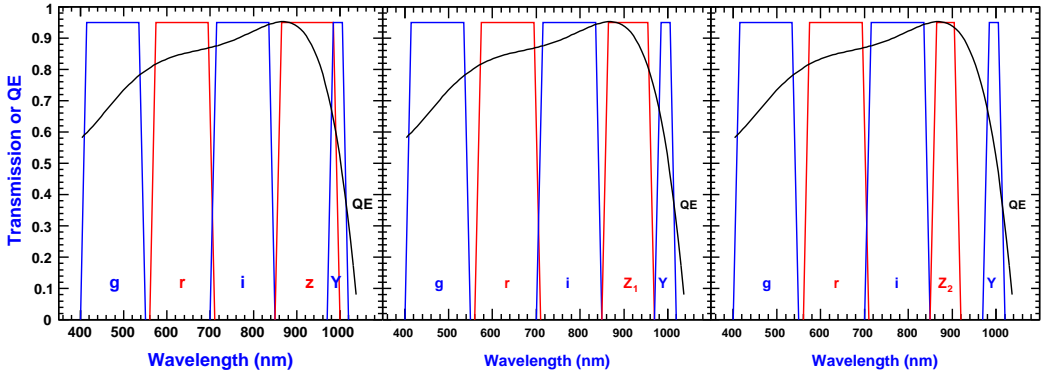}
\caption{The choice of DES z filters plotted with the DES quantum efficiency.}
\label{fig:filt}
\end{center}
\end{figure}

\begin{figure}
\begin{center}
\includegraphics[angle=0,scale=0.8]{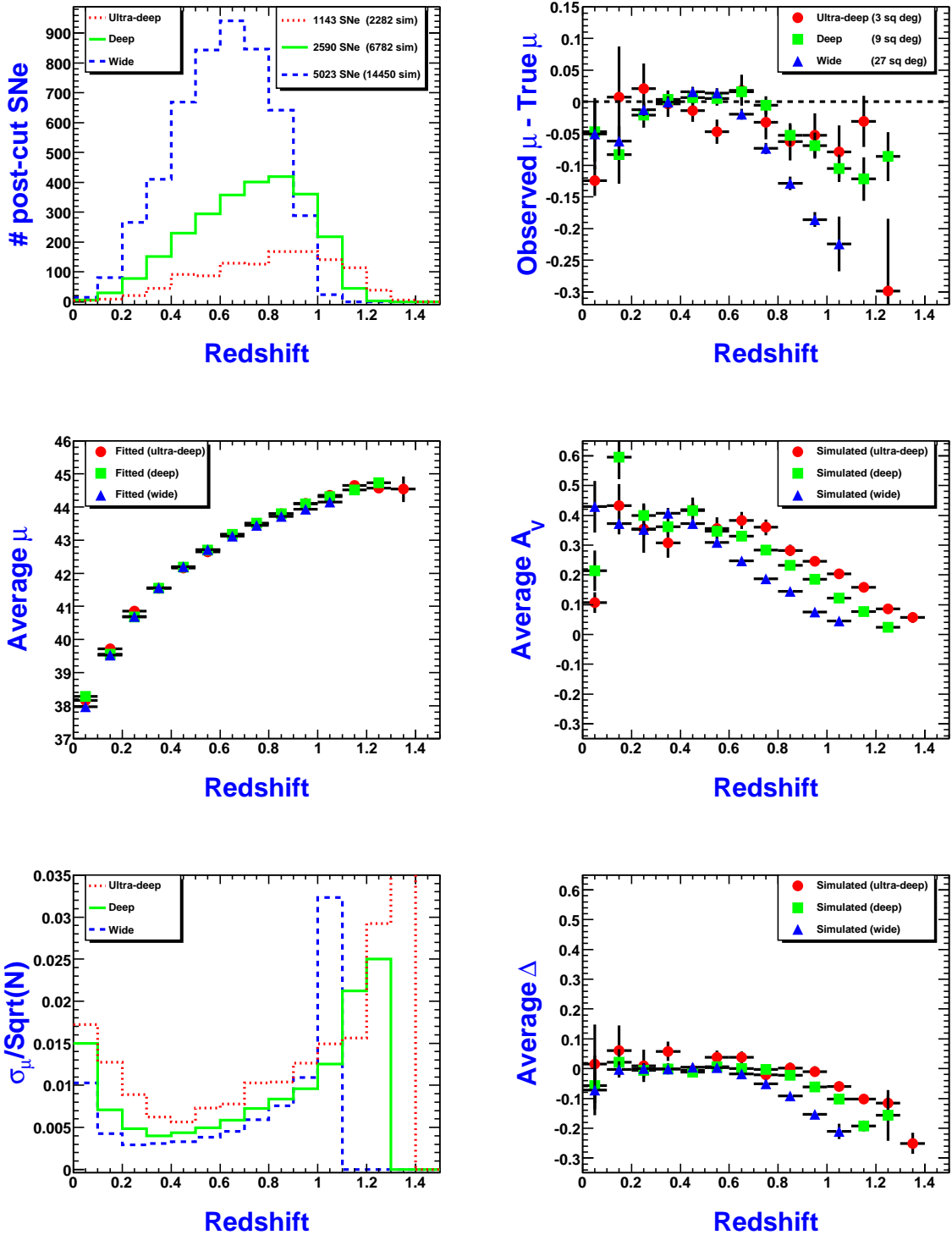}
\caption{\textit{Left}: Number of SNe and the Hubble diagram (grizY filter set). \textit{Right}: The redshift run of the difference in fitted (``observed'') and simulated (``true'') distance modulus ($\mu$), host extinction parameter (A$_{\rm V}$), and MLCS luminosity/shape parameter ($\rm\Delta$). \textit{Both}: cuts were applied by the fitter such that each SN had at least 5 measurements and one filter measurement with a signal to noise above 10 and any 3 filters above 5. Note that the large error bars and deviations at the lowest (\textit{z}$<$0.1) and highest (\textit{z}$>$1.2) redshifts are due to low statistics.
}
\label{fig:sne}
\end{center}
\end{figure}

The DES supernova working group has begun optimizing DES SN survey strategy by exploring the 1) choice of z-like filter and 2) the survey depth. Under consideration are the griz, griZ$_1$Y, and griZ$_2$Y filter sets (see Fig.~\ref{fig:filt}). The griz filters are SDSS-like and the Y filter occupies the clean wavelength range between the atmospheric absorption bands at 0.95$\mu$m and 1.14$\mu$m. Z$_1$ avoids the overlap with Y and Z$_2$ avoids the Y overlap \textit{and} the lower atmospheric absorption feature. Also under consideration are 3, 9, and 27 square-degree fields corresponding to ``ultra-deep'' (but narrow = 1 DES field), ``deep'', and ``wide'' (but shallow) surveys. Results to date show that the survey depth has a much greater effect than does the choice of z-like filter. Therefore, we henceforth show examples using the z filter. Fig.~\ref{fig:sne} shows our DES light curve fits.

\section{Discussion}\label{sec:disc}
Fig.~\ref{fig:sne} shows a $\mu$ bias manifest in the difference between fitted and simulated $\mu$ beyond \textit{z}$\sim$0.6. The bias arises from not accounting for selection efficiencies and illustrates the magnitude of the $\mu$-correction that will be needed. The fact that A$_{\rm V}$ trends to zero beyond \textit{z}$\sim$0.6 is consistent with a selection bias as we interpret that to mean that only less extincted SNe can pass the cuts as redshift increases. Fig.~\ref{fig:sne} also shows that the deep survey offers a substantial improvement in statistics relative to the ultra-deep survey while avoiding a significant portion of the bias suffered by the wide survey. Thus, we will move forward in constraining DES SN strategy by considering both a deep survey and a hybrid approach with a mixture deep and wide fields.

Calculations made for the DES project proposal estimated that the survey would offer an improvement in the the Dark Energy Task Force figure of merit (fom) by a factor of 4.6 relative to current SN surveys. The DES SN working group has implemented a cosmology fitter in order to obtain a more robust calculation of the fom for DES by harnessing SNANA simulated SN surveys. We currently have statistics-only fom estimates and are working on furthering our SNANA analysis to account for estimates of DES SN systematics. Once completed, we will use SNANA to constrain the optimal DES SN survey strategy and produce a detailed white paper.


\section*{References}
\begin{thebibliography}{99}
\bibitem{jha07} S. Jha, A.~G. Riess, R.~P. Kirshner, \Journal{\APJ}{659}{122}{2007}.
\bibitem{guy07} J. Guy {\it et al}, \Journal{\AA}{466}{11}{2007}.
\bibitem{ccm89} J.~A. Cardelli, G.~.C. Clayton, J.~.S. Mathis, \Journal{\APJ}{345}{245}{1989}.
\end{thebibliography}
\end{document}